# NaFe$_{0.56}$Cu$_{0.44}$As: A pnictide insulating phase induced by on-site Coulomb interaction

C. E. Matt[1,2,§], N. Xu[1,3], Baiqing Lv[1,4], Junzhang Ma [1,4], F. Bisti[1], J. Park[1], T. Shang[1,3,5], Chongde Cao[6,7], Yu Song[6], Andriy H. Nevidomskyy[6], Pengcheng Dai[6], L. Patthey[1], N.C. Plumb[1], M. Radovic[1], J. Mesot,[1,2,3] and M. Shi[1,§]

[1] *Swiss Light Source, Paul Scherrer Institut, CH-5232 Villigen PSI, Switzerland*
[2] *Laboratory for Solid State Physics, ETH Zürich, CH-8093 Zürich, Switzerland*
[3] *Institute of Condensed Matter Physics, École Polytechnique Fédérale de Lausanne, CH-10 15 Lausanne, Switzerland*
[4] *Beijing National Laboratory for Condensed Matter Physics and Institute of Physics, Chinese Academy of Sciences, Beijing 100190, China*
[5] *Laboratory for Developments and Methods, Paul Scherrer Institut, CH-5232 Villigen, Switzerland*
[6] *Department of Physics and Astronomy, Rice University, Houston, Texas 77005, USA*
[7] *Department of Applied Physics, Northwestern Polytechnical University, Xian 710072, China*

§ E-mail: christian.matt@psi.ch  ming.shi@psi.ch

In the studies of iron-pnictides, a key question is whether their bad-metal state from which the superconductivity emerges lies in close proximity with a magnetically ordered insulating phase. Recently it was found that at low temperatures, the heavily Cu-doped NaFe$_{1-x}$Cu$_x$As ($x >$ 0.3) iron-pnictide is an insulator with long-range antiferromagnetic order, similar to the parent compound of cuprates but distinct from all other iron-pnictides. Using angle-resolved photoemission spectroscopy, we determined the momentum-resolved electronic structure of NaFe$_{1-x}$Cu$_x$As ($x$ = 0.44) and identified that its ground state is a narrow-gap insulator. Combining the experimental results with density functional theory (DFT) and DFT+U calculations, our analysis reveals that the on-site Coulombic (Hubbard) and Hund's coupling energies play crucial roles in formation of the band gap about the chemical potential. We propose that at finite temperatures charge carriers are thermally excited from the Cu-As-like valence band into the conduction band, which is of Fe 3$d$-like character. With increasing temperature, the number of electrons in the conduction band becomes larger and the hopping energy between Fe sites increases, and finally the long-range antiferromagnetic order is destroyed at $T > T_N$. Our study provides a basis for investigating the evolution of the electronic structure of a Mott insulator transforming into a bad metallic phase, and eventually forming a superconducting state in iron-pnictides.

Similar to the cuprates and iron-chalcogenides, in the phase diagram of the iron-pnictides, a superconducting dome develops upon doping a non-superconducting, often magnetically ordered, parent compound. The superconducting dome and/or magnetic phase are formed from an underlying normal state which exhibits bad-metal behavior with large electrical resistivity at room temperature [1–4]. It has been proposed that the proximity to a Mott insulating phase is responsible for the bad metal behavior and increase in the electronic correlations in those compounds [3,5]. In the studies of iron-chalcogenides, it has also been suggested that the (orbital-dependent) Mott physics plays a key role in the mechanism for the insulating behavior. In cuprates, electron-electron correlations are reduced by doping either electrons or holes into the Mott insulating parent compound, with superconductivity occurring at a few percent of doping [6–8]. In contrast, correlation effects in the iron-pnictides are only weakened with the electron doping, but enhanced when doping holes into their parent compounds [9–12]. However, for all the iron-pnictides, even up to the highest possible hole-doping level (e.g. fully replacing Ba with K in the widely studied Ba$_{1-x}$K$_x$Fe$_2$As$_2$), an antiferromagnetically (AFM) ordered insulating state does not occur [11,13]. It would be highly interesting to identify and investigate an iron-pnictide family that undergoes an AFM insulating – (bad) metallic – superconducting phase transition tuned by doping, and study these compounds' electronic structures in distinct phases.

Very recently it has been shown that, adjacent to the superconducting phase, an AFM insulating state occurs in heavily doped NaFe$_{1-x}$Cu$_x$As [14,15]. For $x > 0.3$, below $T_N$ = 200K the system develops antiferromagnetic order which becomes long-range for $x \geq 0.44$. The spin arrangement is depicted in Fig 1 c), with a magnetic moment $\sim 1.1\mu_B$ per iron site, ten times larger than that in the parent compound NaFeAs. Using angle-resolved photoemission spectroscopy (ARPES), combined with density functional theory (DFT) and DFT+U calculations, we reveal that in NaFe$_{1-x}$Cu$_x$As with high $x$ values ($x = 0.44$), the Hubbard $U$ and Hund's coupling ($J_H$) have a strong effect on the underlying electronic structure, which results in a finite energy gap occurring at the Fermi level ($E_F$). The ARPES results are significantly different from the electronic structure obtained from DFT calculations, but can be reproduced by DFT+U calculations with the inclusion of $U$ and $J_H$ to account for the on-site Coulomb interaction. We show that the main effect of the interactions is to shift the Fe $d$ orbital-related bands to higher binding energies, which results in the ground state of NaFe$_{0.56}$Cu$_{0.44}$As being an insulator.

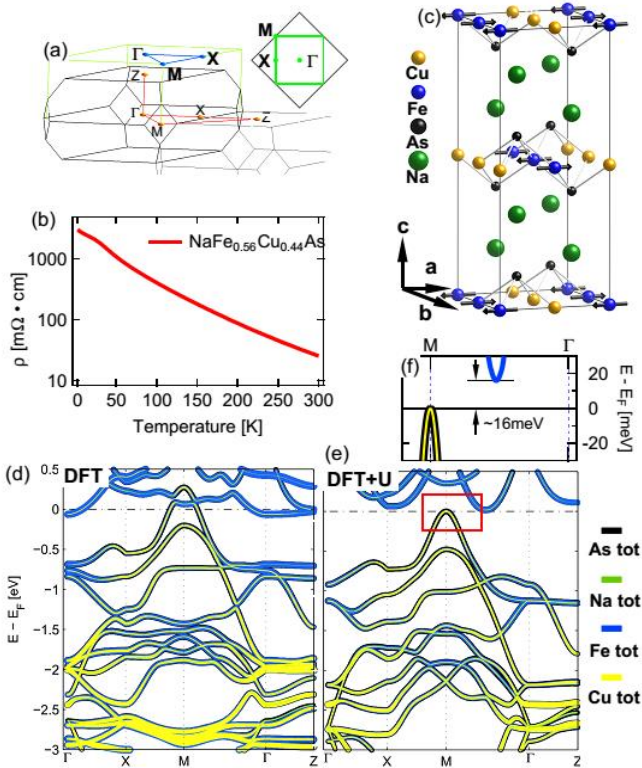

Figure 1: (Color online) Electronic structure of NaFe$_{0.5}$Cu$_{0.5}$As calculated by DFT and DFT+U. (a) Brillouin zone (BZ) with high symmetry points and lines indicated. The light green line indicates the projected BZ. Inset: Light green lines denote the projected BZ of NaFe$_{0.56}$Cu$_{0.44}$As; black lines the BZ of the 2-Fe unit-cell of BaFe$_2$As$_2$ as defined in [16,17]. (b) Temperature dependence of the in-plane resistivity for NaFe$_{0.56}$Cu$_{0.44}$As. (c) The crystal structure [18]. (d) and (e) (h) and (i) The band structure along high symmetry lines from DFT and DFT+U calculations, respectively. (k) Zoom-in of the red box in (i), showing the bandgap about $E_F$.

High-quality single crystals of NaFe$_{1-x}$Cu$_x$As were grown using the self-flux method as described in [15].

ARPES experiments were carried out at the Surface and Interface Spectroscopy (SIS) beamline [19], Swiss Light Source (SLS), using a Scienta R4000 electron analyzer with total energy and angular resolutions of ~20 meV and ~0.15°, respectively. Samples were cleaved in-situ under ultrahigh vacuum (UHV) conditions ($< 5 \times 10^{-11}$ mbar). The DFT calculations were performed using the WIEN2K package [20] with the crystal structure shown in Fig. 1a and the lattice parameters ($a=b=$ 5.72Å, $c=$13.85Å) determined from neutron diffraction on NaFe$_{0.56}$Cu$_{0.44}$As [14] and with interaction parameters $U = 3.15$ eV and $J_H = 0.4$ eV [21].

To get more insight into the insulating behavior of NaFe$_{0.56}$Cu$_{0.44}$As at low temperatures as observed in the transport measurements (Fig 1 b) and [14,15]), we have performed electronic structure calculations without and with on-site correlations by using DFT and DFT+U methods, respectively. The DFT calculation predicts a semi-metallic state with an electron-like pocket at $\Gamma$ (Fe character) and a hole-like pocket around $M$ (Cu-As character); see Fig. 1d). These pockets produce a finite density of states (DOS) at $E_F$ (Figure S1), which is inconsistent with the insulating behavior of NaFe$_{0.56}$Cu$_{0.44}$As. On the other hand, if $U$ and $J_H$ are included in the calculations (DFT+U), the valence bands get pushed to higher binding energies and the conduction bands are shifted up, resulting in a finite energy gap between valence and conduction bands (Fig. 1 e,f). The magnitude of the band shift depends strongly on the elemental and orbital composition of the bands; it is negligible for the Cu- and As-derived bands but large for the bands formed by Fe $d$ orbitals. Among the Fe $d$ orbitals, the on-site Coulomb interaction has the strongest effect on $d_{xy}$-and $d_{z^2}$-orbital dominated energy bands. Similar to the case of BaCu$_2$As$_2$ [22,23], the DFT-band near $E_F$ around the $M$ point is formed by the hybridization of As 4$p$ and Cu 3$d$ orbitals with Fe 3$d$ orbitals [21]. The inclusion of on-site Coulomb interaction lifts the hybridization, pushing the Cu-As band below $E_F$, and shifting the Fe band above $E_F$.

Figure 2 shows ARPES spectra and their curvature plots along the high symmetry lines in the Brillouin Zone (BZ), see Fig 1a). The overlaid lines are the calculated band dispersion renormalized by a factor of 1.35. The overall agreement between the ARPES spectra and the calculated DFT+U electronic structure is significant: (1) no electron-like pocket around $\Gamma$ point was observed near $E_F$, which is consistent with the predication by DFT+U calculations and different from that of DFT; (2) moving from the $\Gamma$ or $X$ point to the $M$ point, the band approaches but does not cross the Fermi level, thus no hole-like pocket around M point is formed and (3), at high binding energies, the ARPES results agree better with DFT+U than with DFT.

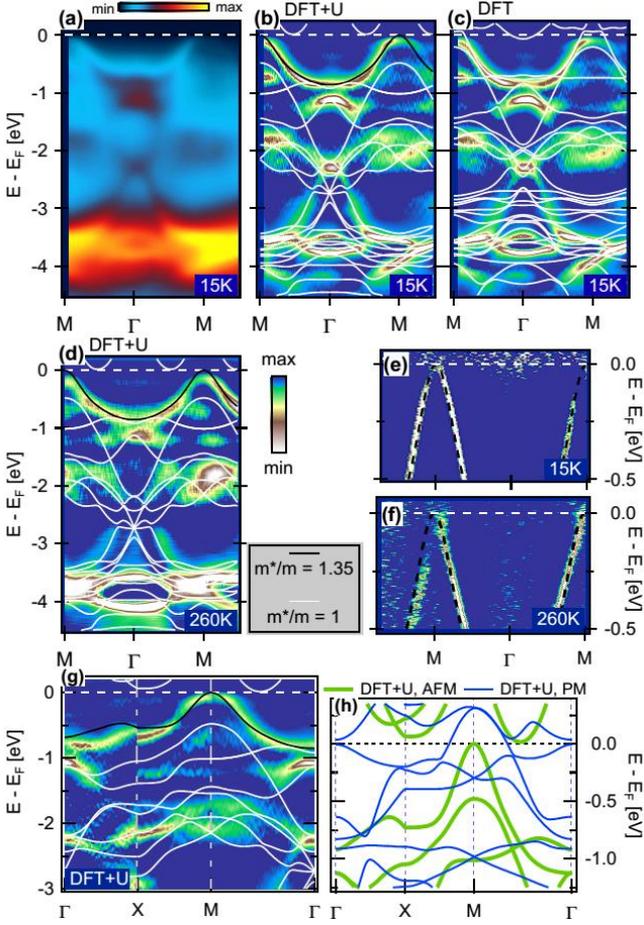

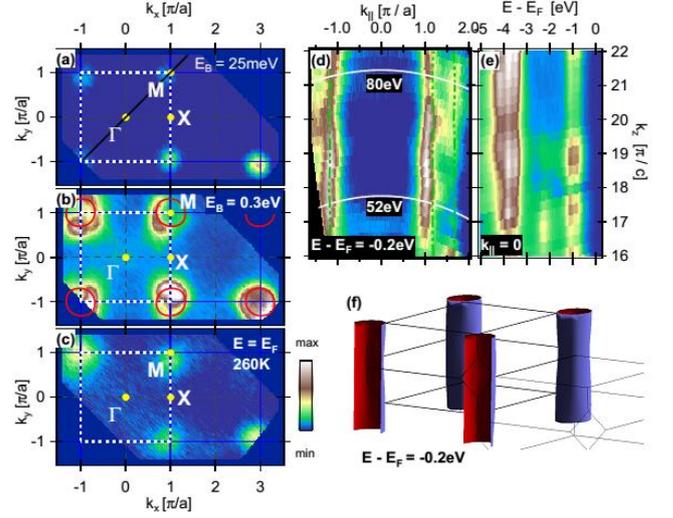

Figure 2. (Color online) ARPES spectra of NaFe$_{0.56}$Cu$_{0.44}$As. (a) ARPES spectrum along M-Γ-M direction, taken at $T$ = 15 K with circularly polarized light and photon energy $h\nu$ = 52 eV. (b) and (c) Curvature plot of the ARPES spectrum in (a), the superimposed lines are the band dispersion from DFT and DFT+U calculations, respectively. The calculated dispersion indicated by the black line is renormalized by a factor of 1.35. (d) The same as (b) but the ARPES spectrum is taken at $T$ = 260 K. (e) and (f) the same as (b) and (d) but close to $E_F$. The black dashed lines indicate the band dispersion from DFT+U calculation. (g) Curvature plot of the ARPES spectrum at $T$ = 15 $K$ along high symmetry lines as indicated in Fig 1 a). The superimposed lines are the band dispersion from DFT+U calculation. (h) Band dispersion as calculated by DFT+U for the AFM (green lines) and paramagnetic state (blue lines).

For example, the multiple flat Fe/Cu-derived bands at ~ 3 eV in the DFT calculations were not observed by ARPES (Fig. 2 c)). We would like to point out that in contrast to the DFT calculations on the majority of iron pnictides, no shift of individual bands is required in order to match the ARPES results. Our ARPES results reveal that the ground state of NaFe$_{0.56}$Cu$_{0.44}$As is an insulator, consistent with the scanning tunneling microscopy (STM) study of heavily Cu-doped NaFe$_{1-x}$Cu$_x$As ($x$ = 0.3), which showed diminished density of states at the chemical potential [24]. In Fig. 2d, f) we plot the ARPES spectra taken at 260 $K$, well above $T_N$ ~ 200 K; the dispersion is almost identical to that at low temperature (15 $K$). We would like to emphasize that the local magnetic moment on the Fe sites plays an essential role for the observed electronic structure below and above $T_N$. As shown in Fig. 2h the band dispersion near $E_F$, obtained from the DFT+U calculation in the paramagnetic state, forms a large, hole-like Fermi surface around M, which is not observed in the ARPES measurement.

Figure 3. (Color online) (a) and (b) ARPES intensity maps at energies $E - E_F$ = -25 meV and $E - E_F$ = -300 meV, respectively. The spectra were acquired with 52 eV circularly polarized light at $T$ = 15 K. The maps have been integrated over an energy window of ±10 meV. (c) ARPES intensity map at $E_F$, obtained at $T$ = 260 K. Due to the thermal broadening, finite intensity appears at the $E_F$. (d) ARPES intensity map at energies $E - E_F$ = - 200 meV in the $k_{\parallel}$ - $k_z$ plane, where $k_{\parallel}$ is along Γ-M direction. (e) The dispersion along $k_z$ direction at $k_{\parallel}$ = 0. (f) Sketch of the constant energy surface at $E - E_F$ = -200 meV in 3D BZ, calculated by DFT+U method. (g) Schematic of the electronic structures of cuprate parent compounds and NaFe$_{0.56}$Cu$_{0.44}$As. The quantity ~$k_B T$ indicates the thermal excitations of electrons from the valence to conduction bands.

To further explore the insulating or semi-metallic behavior of NaFe$_{1-x}$Cu$_x$As, we have carried out ARPES measurements in several Brillouin zones. Figure 3 shows ARPES intensity maps at fixed binding energies. In Fig. 3a, b), the intensity map obtained at 15K is plotted 25 meV and 300 meV below $E_F$ since the spectral intensity at $E_F$ is vanishing. As expected, except for the (Cu, As)-formed valence band (Fig. 2a-b,e,g) predicted by the DFT+U calculation, no other band appears at these two energies. The ARPES intensity maps at different binding energies vary little with $k_z$, indicating the electronic structure is quasi two-dimensional in NaFe$_{0.56}$Cu$_{0.44}$As. These observations are in concert with DFT+U calculations and provide spectroscopic evidence that the system is insulating at low temperatures. In the DFT+U calculations the effect of including $U$ and $J_H$ is to remove the hybridization of As *p* and Cu *d* with Fe 3*d* orbitals and push the bands dominated by Fe 3*d* orbitals away from the chemical potential. The remaining band near $E_F$ is mainly composed by Cu orbitals (Fig. 1 e). It has been shown in ref. [14] that the Cu is close to $d^{10}$ configuration. This could explain why the valence band near $E_F$ is only renormalized by a factor of ~ 1.35 (Fig. 2e), which is much smaller than the renormalization factor in Fe-pnictide parent or hole-doped superconducting Fe-pnictide compounds (typically ~ 2 – 4) [11,25–27] whose Fe 3*d* orbitals are partially filled.

In Fig. 3c, we plot the ARPES intensity map at $E_F$ at $T = 260$ K. Due to thermal broadening, we observe finite spectral weight around *M*, very similar to the intensity map taken at low temperatures, close to $E_F$ (see Fig. 3a). The primary results of our ARPES measurements are the observation of a valence band which is touching, but not crossing the Fermi level. Surprisingly, the observed band dispersion is highly temperature independent and strongly resembles the low-temperature results for temperatures far above $T_N$.

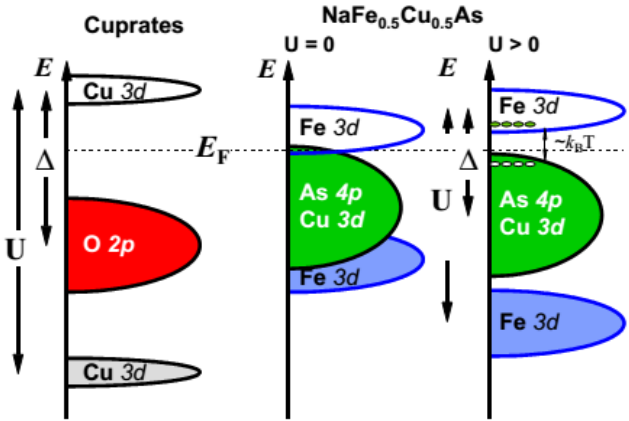

Figure 4: Schematic of the electronic structure of cuprate parent compounds and NaFe$_{0.56}$Cu$_{0.44}$As for the noninteracting case ($U = 0$) and or the case of finite interactions. The quantity ~$k_BT$ indicates the thermal excitations of electrons from the valence to conduction bands.

Having established that NaFe$_{0.56}$Cu$_{0.44}$As is an insulator in the ground state due to correlation effects, we turn to the natural question of how the long-range AFM order disappears at temperatures above Néel ordering temperature ($T_N \sim 200$ K), as observed in neutron scattering measurements [14]. In Mott insulators the AFM ordering of local magnetic moments is related to the superexchange energy, which is proportional to $t^2/U$, where $t$ is the hopping integral that scales with the kinetic energy of charge carriers moving in solids [6,28,29]. If the on-site Coulomb repulsion exceeds the kinetic energy, the intersite hopping of charge carriers is blocked, causing them to become localized. On the other hand, if the Coulomb energy barrier can be overcome due to a large hopping energy, electrons or holes are delocalized and the long-range AFM ordering is suppressed. It is known that long-range AFM order in a Mott insulator can be suppressed by changing the band filling with chemical doping. For example, in cuprates, at low temperatures, long-range AFM order is destroyed by doping a few percent of holes or electrons into the Mott insulating parent compounds. Note that the Mott gap (or more generally, charge transfer gap) needs not to be suppressed completely for the long-range AFM order to disappear: the insulating Mott state without long-range magnetic order often persists above the Néel temperature [30–32]. This consideration provides a possible explanation for the AFM ordering in NaFe$_{0.56}$Cu$_{0.44}$As observed in neutron scattering experiments, as well as its temperature-dependence. Figure 4 schematically depicts the DOS of NaFe$_{0.56}$Cu$_{0.44}$As in comparison with the parent compounds of cuprates [33]. Due to the on-site Coulombic repulsion, the Fe *d* orbital-related conduction and valence bands move farther away from the chemical potential (Fig. 4). About $E_F$, a small indirect band gap occurs between the top of the valence band formed by (Cu, As) and the bottom of conduction bands dominated by Fe *d* states. In the ground state, the system is insulating: the virtual hopping of electrons with antiparallel spins from one Fe site to the next, restricted by the Pauli exclusion principle, leads to the formation of an AFM arrangement in the Fe lattice. Upon increasing temperature, due to the small band gap and thermal excitation, the lowest conduction band starts to be populated by electrons from the highest Cu–As valence band, which leaves holes in the valence band (see Fig 4, right panel). Because the holes (electrons) in the valence (conduction) band are mobile, the resistivity of the material decreases, as manifested in transport measurements (see Fig 1 b), similarly to thermally activated transport in a narrow-band semiconductors. This hopping of the thermally excited electrons in the lowest conduction band (which has Fe 3*d* character, see Fig. 1 e,f) suppresses the long-range AFM order in the Fe lattice, while the charge gap persists. This is consistent with the non-magnetic insulating behavior of resistivity observed well above the Néel temperature in NaFe$_{1-x}$Cu$_x$As (see Fig 1 b) [14,15]. Our ARPES results obtained at 260 *K* (well above $T_N$) show that the band dispersion as well as the band renormalization near $E_F$ are very similar to that at 15 *K* (well below $T_N$) (Fig. 2 c-f), indicating that there is no significant change in the electronic structure as $T_N$ is crossed. This would suggest that the disappearance of the long-range AFM ordering at T > 200 K is indeed related to the occupation of the lowest conduction and highest valence bands, without closing the Mott charge-transfer gap. The robustness of the electronic structure is consistent with transport measurements: upon increasing temperature the resistivity smoothly decreases and the insulating behavior persists at least up to 300 $K$; no abnormal behavior is observed in the vicinity of $T_N$; see Fig 1 b). While not excluding other more sophisticated models (e.g., orbital-selective Mott physics [9,10,34]) that would be able to quantitatively account for the experimental results on this material, the basic picture described here provides a qualitatively consistent explanation for the experimental results from neutron scattering, transport and ARPES measurements. This picture in which Mott physics plays a key role for the insulating behavior in highly Cu-doped NaFe$_{1-x}$Cu$_x$As is strongly supported by a recent STM study on insulating NaFe$_{0.7}$Cu$_{0.3}$As which reports striking similarities to lightly doped cuprates [24]. The early ARPES studies on NaFe$_{1-x}$Cu$_x$As with $x$ up to 0.14 showed that, except for introducing extra charge carriers, the overall band dispersion barely changes with Cu doping, and the Fermi surface and/or all the energy bands near $E_F$ are dominated by Fe 3$d$ orbitals [35]. It is of high interest to reveal the momentum-resolved electronic structure with ARPES in order to get further insight into the evolution of the Cu 3$d$-derived electronic states at or near $E_F$, as the long-range AFM-ordered insulator evolves into a metallic or superconducting state with decreasing the concentration of the Cu dopant in NaFe$_{1-x}$Cu$_x$As.

In summary, using ARPES combined with DFT and DFT+U calculations, we have revealed the electronic structure of the heavily Cu-doped NaFe$_{1-x}$Cu$_x$As ($x$ = 0.44) and showed that its ground state is a narrow-gap insulator whose origin lies in strong electron interactions of Fe 3$d$ orbitals. The on-site interaction (Hubbard $U$ and Hund's coupling $J_H$) remove the hybridizations between Fe 3$d$ and other Cu-As orbitals in the vicinity of $E_F$. The interactions furthermore elevate the Fe 3$d$ bands near $E_F$ to above the chemical potential and push the fully occupied Fe 3$d$ valence band further down in binding energy. Consequently, an energy gap opens up about the chemical potential, which instigates an insulating phase in the heavily Cu-doped NaFe$_{1-x}$Cu$_x$As. The effect induced by $U$ and $J_H$ resembles the situation in the parent compounds of the high-$T_c$ cuprates (Mott insulators), with the Mott induced charge transfer gap that underscores the insulating behavior. The quantitative differences with cuprates are that the top of the valence band is derived from Cu–As hybridized bands and is very close to (or touching) the chemical potential; and the charge transfer gap between valence and conduction bands is very narrow. The occupation of Fe 3$d$ conduction band by thermally activated electrons increases the hopping processes between the Fe sites and thus suppresses the AFM ordering, eventually destroying the long-rage order at $T$ > $T_N$.

**Acknowledgements**
We thank T. Schmitt, Zhiming Wang and Bruce Normand for their help and enlighting discussions.
This work was supported by the Swiss National Science Foundation (No. 200021-159678) and the Sino-Swiss Science and Technology Cooperation (project number IZLCZ2138954). The single crystal growth is supported by the U.S. DOE, BES under contract no. de-sc0012311 and by the National Natural Science Foundation of China Grant No. 51471135. The theoretical work at Rice was supported by the Robert A. Welch Foundation Grant No. C-1818 (A.H.N.) and by U.S. NSF grant DMR-1350237 (A.H.N.).